\documentclass[a4paper,11pt]{article}
\usepackage{pos}
\usepackage{physics}
\usepackage{hyperref}

\title{Status of the Short-Baseline Near Detector at Fermilab}

\author*[a]{Rodrigo Alvarez-Garrote \footnote[2]{on behalf of the SBND collaboration}}

\affiliation[a]{Centro de Investigaciones Energ\'eticas, Medioambientales y Tecnol\'ogicas (CIEMAT),\\
  Avenida Complutense 40, 28040 Madrid, Spain}

\emailAdd{rodrigo.alvarez@ciemat.es}

\abstract{

The Short-Baseline Near Detector (SBND) is one of three Liquid Argon Time Projection Chamber (LArTPC) neutrino detectors positioned along the axis of the Booster Neutrino Beam (BNB) at Fermilab, as part of the Short-Baseline Neutrino (SBN) Program. The detector is currently being commissioned and is expected to take neutrino data this year. SBND is characterized by superb imaging capabilities and will record over a million neutrino interactions per year. Thanks to its unique combination of measurement resolution and statistics, SBND will carry out a rich program of neutrino interaction measurements and novel searches for physics beyond the Standard Model (BSM). It will enable the potential of the overall SBN sterile neutrino program by performing a precise characterization of the unoscillated event rate, and constraining BNB flux and neutrino-argon cross-section systematic uncertainties. In this proceedings, the physics reach, current status, and future prospects of SBND are discussed and early data is presented.}

\FullConference{%
  International Conference on High Energy Physics\\
  17-24 July 2024\\
  Prague Congress Centre, Prague, Czech Republic
}

\begin{document}
\maketitle

\section{Introduction}

One of the open questions in neutrino physics is the anomalous results reported by some accelerator and reactor experiments. In particular, the LSND and MiniBooNE collaborations \cite{LSND_1, miniboone_excess} have observed an excess of electron-like events over the past decades. A proposed hypothesis is the addition of a fourth light sterile neutrino mixing with Standard Model neutrinos via an extended PMNS matrix \cite{dasgupta2021sterile}. The Liquid Argon Time Projection Chamber (LArTPC) technology  \cite{lartpc_rubbia} offers an ideal platform for investigating these anomalies, owing to its precise three-dimensional event reconstruction and capabilities to distinguish electron and photon signals. In 2014, the Short-Baseline Neutrino program \cite{sbn_proposal}  was proposed to finally resolve this puzzle.  

\section{The Short-Baseline Neutrino Program}

\begin{figure}[h]
\centering
\includegraphics[width=\linewidth]{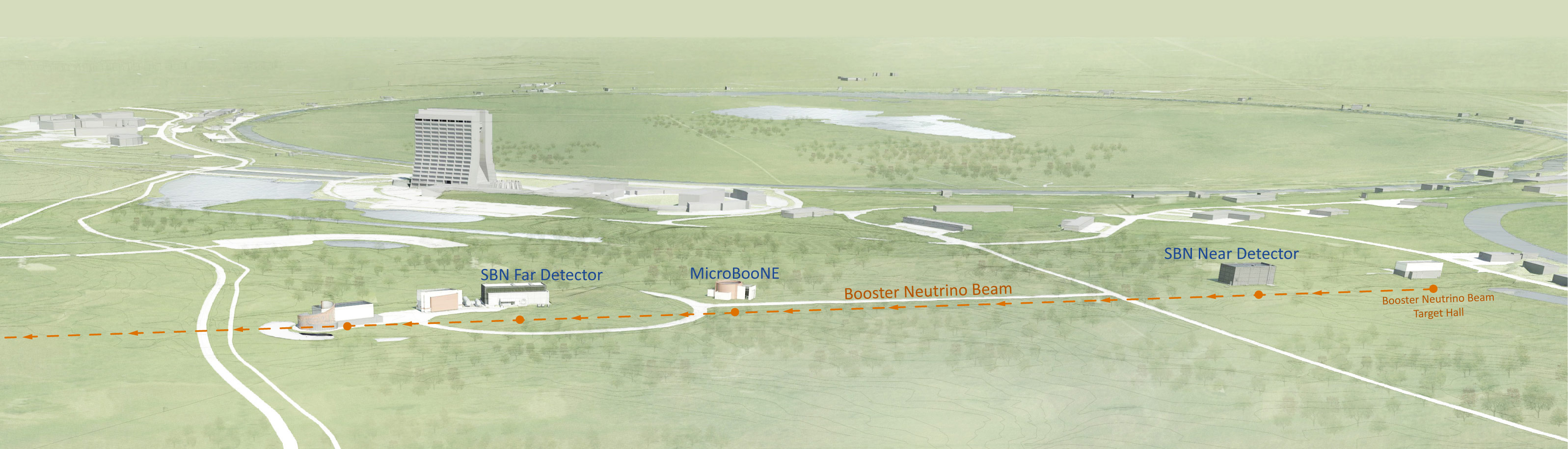}
\caption{Map of the SBN program at Fermilab. The three SBN detectors (in blue): SBND, MicroBooNE and ICARUS, have baselines of 110 m, 470 m and 600 m respectively. The Booster Neutrino Beam (in red) delivers a 99.5\% $\nu_\mu$ flux with $\expval{E_{\nu_\mu}}=0.8$ GeV.}
\label{fig:SBN_map}
\end{figure}

The Short-Baseline Neutrino (SBN) program at Fermilab consist of three LArTPC experiments at different distances from the Booster Neutrino Beam (BNB): the near detector SBND, MicroBooNE (almost at the MiniBooNE location) and ICARUS, the far detector (see Figure \ref{fig:SBN_map}). The SBN program will look for light sterile neutrinos in the eV-scale and test the LSND and MiniBooNE anomalies. To constrain the systematic uncertainties to the percent level, the three detectors employ the same technology and target medium, the LArTPC. 

MicroBooNE, at 470 meters of the BNB target was the first of the three and operated from 2015 to 2021 \cite{microboone_tdr}. ICARUS had been taking data since 2010 at Gran Sasso (Italy) and was moved to Fermilab in 2017. The experiment re-started operations at its new location on 2020 \cite{ICARUS_at_sbn_first_results}. SBND, the near detector of the SBN program, is only 110 m away from the BNB target. 

\section{SBND: the Short-Baseline Near Detector}

The Short-Baseline Near Detector (SBND) is a LArTPC experiment with an active volume of $4\times4\times5$ meters or 112 tons of liquid argon. The detector was filled between February and April and is currently being commissioned. Data taking at the nominal electric field started in July. Physics quality runs will start in fall 2024 as the BNB resumes operations after the summer shutdown. The SBND physics program includes:

\begin{itemize}
    \item The search of eV-scale sterile neutrinos, as a part of the SBN program. Through measuring the un-oscillated neutrino flux, SBND will play a key role to understand the excesses of electron-like events observed by MiniBooNE and LSND.
    \item With O($10^6$) neutrino interactions in 3 years of data taking,  SBND will measure many neutrino-nucleon cross sections on argon with world-leading statistics. Constraining the statistical uncertainties below the percent level for many channels, SBND data will pave the way for next-generation LArTPC experiments.
    \item A broad BSM program \cite{supraja_2023_sbnd_bsm} that includes searches for long-lived particles (heavy neutral leptons, heavy QCD axions...), dark photons...
    \item Research and development in new light detection technologies such as X-ARAPUCA sensors \cite{xarapuca_1st_paper} and TPB-coated reflective foils. This is of special relevance for the DUNE experiment, whose light detection system is fully composed by X-ARAPUCAs.
\end{itemize}

\section{SBND design and status}

\begin{figure}[ht]
    \centering
    \includegraphics[width=0.539\linewidth]{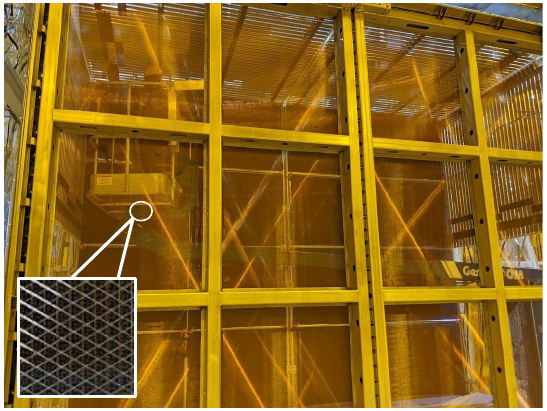}
    \includegraphics[width=0.45\linewidth]{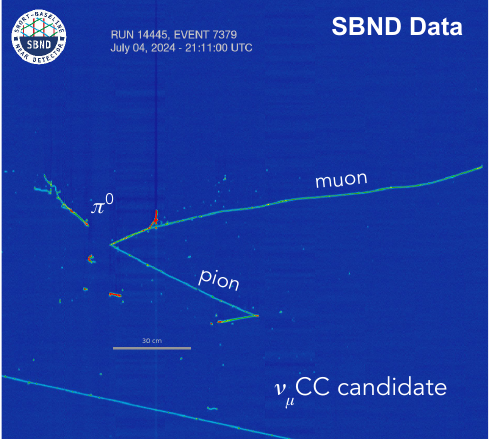}
    \caption{Left: APA assembled at Fermilab and close-up of the wire planes. Right: Charged-current $\nu_\mu$ candidate from July 2024 early data, a cosmic muon background can be seen in the bottom-left corner.}
    \label{fig:TPC}
\end{figure}

SBND's active volume is split in two TPCs by the cathode plane in the middle. The anode plane assemblies (APAs) are 200 cm away each from the cathode. Each anode plane has two induction and one collection plane, for a total 11264 wires with a wire-pitch of 3 mm (see Fig \ref{fig:TPC}). The collection plane has a vertical orientation while the wires in the two induction planes are angled $\pm60^\circ$ to the vertical plane. The cathode is set at a voltage of -100 kV and the anode plane APAs at ground for a nominal field of 500 V/cm. After the TPC high voltage ramp-up in early July, data taking with neutrino beam began. One of the first neutrino candidates recorded is shown in Figure \ref{fig:TPC}. The TPC noise and argon purity have been monitored with values meeting the design requirements.

To mitigate the background from cosmic ray muons, the SBND cryostat is surrounded by a Cosmic Ray Tagger (CRT) system made of plastic scintillator panels. The superposition of consecutive orthogonal planes allows to 3D position the charged particles crossing  the CRT walls. The CRT system installation started in 2019, resumed after the detector filling and finished with the top walls in September 2024.

\begin{figure}[ht]
    \centering
    \includegraphics[width=0.66\linewidth]{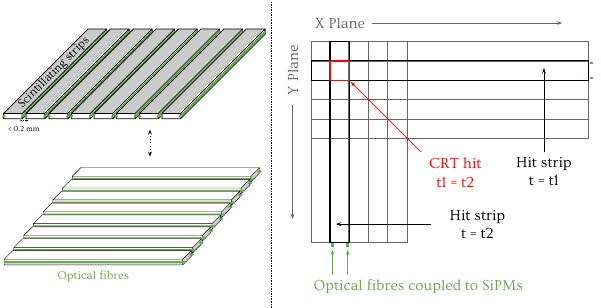}
    \includegraphics[width=0.33\linewidth]{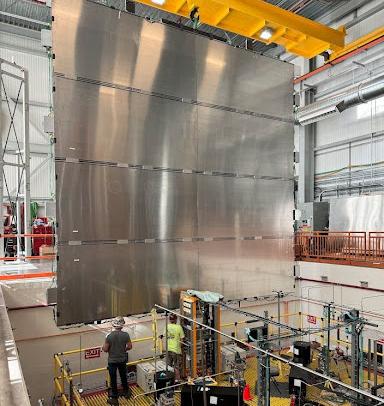}
    \caption{Left and center: scheme of the CRT working principle: the scintillation light is transported with an optical fiber, SiPMs in the strip end capture the photons. Right: installation of the south CRT wall in April 2024.}
    \label{fig:evd}
\end{figure}

\begin{figure}[ht]
    \centering
    \includegraphics[width=\linewidth]{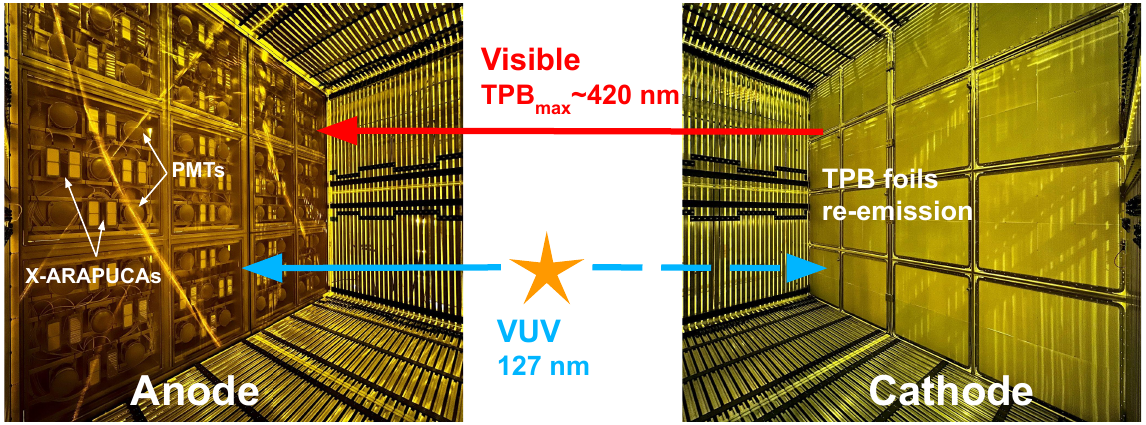}
    \caption{Left: photo of the PDS sensors located behind the TPC wire planes. Right: image of the cathode plane and the TPB-coated reflective foils. PDS sensors collect direct VUV light (blue) and re-emitted photons by the TPB in the visible range (red).}
    \label{fig:PDS}
\end{figure}

SBND incorporates a novel photon detection system (PDS), composed of 120 8-inch Hamamatsu R5912-MOD Cryogenic PMTs and 192 X-ARAPUCA \cite{xarapuca_1st_paper} units (see Figure \ref{fig:PDS}-left). The cathode plane is equipped with Tetraphenyl Butadiene (TPB) coated foils, shown in Figure \ref{fig:PDS}-right, that re-emit the VUV scintillation photons in the visible range. 4 out of each 5 PMTs and half of the X-ARAPUCA units are coated with wavelength shifters (TPB and pTP, respectively). The coated units are sensitive to the direct VUV light and the visible light produced by the foils in the cathode plane while uncoated units are only sensitive to the visible component. The foils recover part of the light yield for depositions away from the anode planes. The combination of units sensitive to light produced in different locations of the detector allows SBND to reconstruct the drift direction of neutrino events using only light information \cite{sbnd_pds_2024}.

\begin{figure}[ht]
    \centering
    \includegraphics[width=0.524\linewidth]{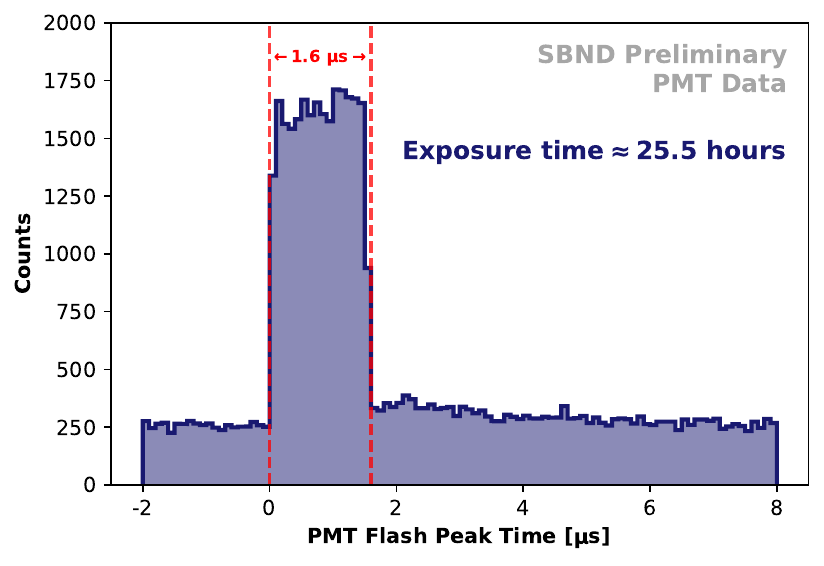}
    \includegraphics[width=0.47\linewidth]{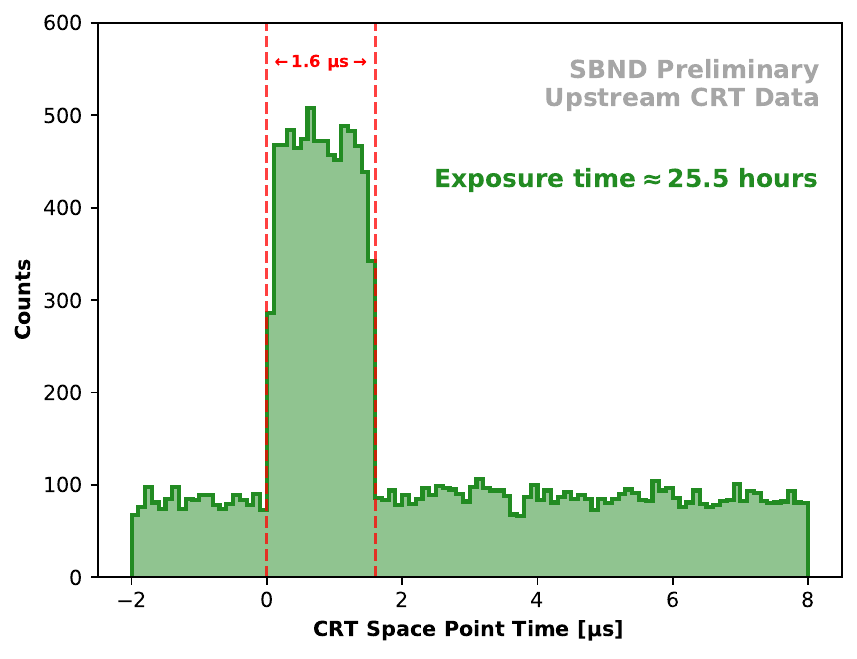}
    \caption{Beam arrival gate plots for clustered PMT signals (flashes) and CRT (space points). The excess from beam-events during the BNB window beam-events (1.6 $\mu$s wide) is clearly visible over the constant cosmic background. The CRT upstream wall is located before the TPC in the beam direction. The excess of events in this wall comes mainly from \textit{dirt} neutrino interactions with the concrete of the building before reaching the detector. }
    \label{fig:top_hat_plots}
\end{figure}

The cosmic muon time distribution is uniform in time, whereas the BNB issues neutrino spills with a width of 1.6 $\mu s$. By associating CRT and PDS signals with the corresponding TPC ones, we can remove the majority of the cosmic backgrounds as they are \textit{out of time} with respect to the beam window. Figure \ref{fig:top_hat_plots} shows the flat contribution from cosmic background signals and the expected excess in the beam window recorded by the PDS(left) and CRT (right) systems.

\section{Summary}

The SBND detector has started operations at nominal conditions in summer 2024. Early data shows the expected excess in the beam window, and the first neutrino candidates have been recorded. The experiment will resume operations in fall 2024 after the BNB summer shutdown. SBND will record millions of neutrino interactions in the following years with world-leading cross-section measurements in argon and a rich BSM program. In conjunction with MicroBooNE and ICARUS, SBND will address the MiniBooNE and LSND anomalies and test the eV-scale sterile neutrino hypothesis with high precision.

\bibliographystyle{JHEP}
\bibliography{skeleton}

\end{document}